\documentclass[twocolumn,aps,superscriptaddress,longbibliography]{revtex4-2}
\usepackage{bm}
\usepackage{amsmath}
\usepackage{amssymb}
\usepackage{times}
\usepackage[usenames,dvipsnames]{xcolor}
\usepackage{graphicx}
\usepackage{dcolumn}
\usepackage{hyperref}
\usepackage{float}
\hypersetup{
  colorlinks=false,
  colorlinks=true,
  citecolor=blue,
  linkcolor=blue,
  urlcolor=blue}

\DeclareUnicodeCharacter{2212}{-}

\begin{document}

\title{Transition metal (group V) doping induced spin and valley polarization in MoS$_2$ monolayer }

\author{Shivani Kumawat}
\affiliation{Department of Physics, Indian Institute of Technology,
             Hauz Khas, New Delhi 110016, India}

%\author{Chandan Kumar Vishwakarma}
%\affiliation{Materials Department, University of California, Santa Barbara, California 93106-5050, USA}

%\author{Mohd Zeeshan}
%\affiliation{Department of Physics, Indian Institute of Technology,
%             Hauz Khas, New Delhi 110016, India}

%\author{Indranil Mal}
%\affiliation{Department of Physics, Indian Institute of Technology,
%             Hauz Khas, New Delhi 110016, India}

\author{Sunil Kumar}
\email{kumarsunil@physics.iitd.ac.in}
\affiliation{Department of Physics, Indian Institute of Technology,
             Hauz Khas, New Delhi 110016, India}

\author{B. K. Mani}
\email{bkmani@physics.iitd.ac.in}
\affiliation{Department of Physics, Indian Institute of Technology,
             Hauz Khas, New Delhi 110016, India}

\begin{abstract}

Doping in two-dimensional materials has emerged as an effective tool for
modulating their electronic properties and thereby enabling their 
multifunctional applications. In this work, we present a first-principles 
study on induced effective magnetic moment and metallicity in MoS$_2$ 
monolayer by substitutional doping of group-5 transition 
metal (TM) elements -- V, Nb and Ta. From our study, we observe that 
the V doping induces half-metallicity, whereas metallic characteristics 
are observed in the case of Nb and Ta doping. 
Moreover, V and Ta-doped MoS$_2$ monolayers are observed to show total 
induced magnetic moments of 0.922 and 0.624 $\mu_{\rm B}$, respectively. 
Importantly, the combined effects of strong spin-orbit coupling (SOC), broken 
inversion symmetry, and structural asymmetry is observed to lead to a 
permanent valley polarization in the V- and Ta-MoS$_2$ systems. 
In particular, we observed a valley polarization of 121 and 21 meVs for 
V and Ta-doped MoS$_2$, respectively.
Furthermore, an enhanced piezoelectric coefficient for the doped systems is 
observed compared to pristine MoS$_2$. Notably, the simultaneous presence 
of half-metallicity, substantial valley polarization, and enhanced 
piezoelectricity in V-doped MoS$_2$ establishes this system as a promising 
multifunctional platform for next-generation spintronic, valleytronic, 
and piezoelectric nanodevices. Overall, our findings provide fundamental
insights into engineering coupled spin-valley-mechanical degrees of 
freedom in two-dimensional materials for advanced quantum and
nanoelectronic applications.

\end{abstract}

\date{\today}

\maketitle

\section{Introduction}

In recent years, the search for versatile two-dimensional (2D) materials 
which can be useful in multifunctional applications has emerged as a key
idea in materials science. In this direction, monolayer transition
metal dichalcogenides (TMDs), specifically MoS$_2$, have gained 
significant attention. Quantum confinement and reduced dimensionality of 
the MoS$_2$ monolayer lead to its promising electronic, optical and mechanical 
properties \cite{LI201533, burse2025unleashing}. However, it is intrinsically 
non-magnetic, which limits its potential for emerging fields which require 
simultaneous control over spin and valley degrees of freedom.

To overcome this challenge, researchers have tried various approaches 
to induce magnetism in non-magnetic materials. One effective method is 
by doping with transition metal (TM) atoms, which not only introduces a 
magnetic moment but also enhances electromechanical coupling and lifts 
the valley degree of freedom through symmetry breaking. This approach 
has opened up new possibilities for developing ultra-thin spintronic 
devices with enhanced functionality. Theoretically, induced magnetism 
is reported in Mn and Fe-doped MoS$_2$ monolayer \cite{ramasubramaniam2013}, 
and other 3d transition metal doping (V, Cr, Mn, Fe, Co) in monolayer 
MoS$_2$ \cite{mishra2013}. Apart from TMDs, other 2D materials such as 
graphene \cite{yazyev2007,chan2008}, hexagonal boron nitride \cite{huang2014}, 
and black phosphorus \cite{hashmi2015} also show induced magnetism by 
TM doping. In terms of experimental study, V-doped WSe$_2$ monolayer 
was successfully synthesized and shows room-temperature ferromagnetism 
\cite{cheng2013}. A recent study demonstrated tunable magnetism in 
Fe-doped MoS$_2$ monolayers \cite{zhang2015} and dopant concentration 
dependent magnetism \cite{ma2012}.

Apart from spintronics, 2D materials have also shown captivating
valley-related physics, which is an emerging field that utilizes the
valley degree of freedom as an information carrier 
\cite{schaibley2016,mak2018}. It leads to multifunctionality in 
practical applications such as valley Hall effect transistors 
\cite{xiao2012}, valley-selective optics \cite{mak2014}, valleytronic 
logic gates \cite{vitale2018}, and valley-polarized light-emitting 
devices \cite{ye2016}. Among various 2D materials, monolayer TMDs are 
promising candidates for valleytronics. They possess broken inversion 
symmetry, which is crucial to enable valley for practical purposes.
Along with the heavy transition metal atoms in these materials, 
they contribute to strong spin-orbit coupling, which makes TMDs  
an ideal material for exploring valleytronics. However, the limitation 
associated with pristine TMD monolayers is that their valleys remain 
degenerate in energy. Therefore, to break the valley degeneracy is 
the major challenge to create a stable valley polarization. For 
practical valley-based devices, it is required to achieve controllable 
valley polarization, where one valley is preferentially occupied 
over the other. This can be accomplished by lifting the energy 
degeneracy between inequivalent valleys, typically located at 
$K$ and $K'$ points in the hexagonal Brillouin zone. There are 
various mechanisms to induce valley polarization which include 
optical pumping with circularly polarized light \cite{cao2012,zeng2012}, 
using external electric fields \cite{li2014}, or transition metal 
doping \cite{cheng2013valley}. Among various methods, introducing 
magnetic ordering through transition metal doping is an effective 
strategy. The magnetic exchange interaction breaks the valley 
degeneracy and can lead to spontaneous valley polarization. Previous 
theoretical studies have predicted valley splitting in TM-doped TMDs 
and Janus TMDs \cite{zhao2017,zhang2019valley, zhang2017janus, dong2019}.

Additionally, 2D TMDs MoS$_2$ also exhibit piezoelectricity due 
to their non-centrosymmetric crystal structure \cite{wu2014piezoelectricity, 
zhu2015observation}. Piezoelectric materials generate electric polarization
when mechanical stress is applied, or conversely, they deform when an electric 
field is applied. This property makes them useful for various applications, 
including sensors, actuators, and energy harvesting devices. 
The piezoelectric response in 2D materials is particularly interesting 
because it can be much stronger than their bulk counterparts due to reduced 
screening effects and increased flexibility at the nanoscale 
\cite{blonsky2015ab,hinchet2018piezoelectric}.

When piezoelectricity coexists with magnetic and valley properties in the 
same material, it creates opportunities for multifunctional devices where 
mechanical, electronic, magnetic, and optical responses can be coupled together. 
Such materials could enable new types of devices that respond to multiple 
external stimuli simultaneously. The present work aims to explore
and propose a material which could offer spintronic, valleytronic 
and piezoelectric properties simultaneously. By using first-principles 
calculations, we systematically investigate the electronic, magnetic, 
spintronic, valleytronic, and piezoelectric properties of the group 5 
transition-metal (TM) (V, Nb, and Ta) doped MoS$_2$ monolayer. We address 
the following key features: 
(i) the electronic structure of pristine and TM-doped MoS$_2$ 
monolayer; 
(ii) emergence of half-metallicity and induced magnetism after 
TM substitution; (iii) induced valley polarization by V/Ta doping; 
and (iv) enhancement of piezoelectricity compared to pristine MoS$_2$.
Our study provides useful insights into the design of 2D materials that 
can simultaneously utilize spin, valley, and mechanical degrees of freedom 
for advanced technological applications.

The paper is arranged into four sections. In Sec. II, we provide a brief
description of the computational methods and parameters used in our
calculations. In Sec. III, we present and analyze our results on
electronic structure, magnetic properties, valley polarization 
and piezoelectric properties in TM-doped MoS$_2$. The summary of our 
results is presented in the last section of the paper.

%%%%%%%%%%%%%%%%%%%%%%%%%%%%%%%%%%%%%%%%%%%%%%%%%%%%%%%%%%%%%%%%%%%%%%%
%%%                       Method employed                         %%%%
%%%%%%%%%%%%%%%%%%%%%%%%%%%%%%%%%%%%%%%%%%%%%%%%%%%%%%%%%%%%%%%%%%%%%%%
\section{ Computational Methodology}

We performed density functional theory (DFT) based first-principles
calculations by using the Vienna Ab initio Simulation Package (VASP)
\cite{KRESSE199615, PhysRevB.54.11169}. To account for the interaction
among electrons, we used the generalized gradient approximation (GGA)-based
Perdew-Burke-Ernzerhof (PBE) \cite{PhysRevLett.77.3865} exchange-correlation
functional. We used the projector augmented wave (PAW) method
\cite{PhysRevB.50.17953} based pseudopotential to accurately represent
the wave functions of valence electrons near the atomic cores. An
energy cutoff of 450 eV was used as a plane-wave basis set, whereas
the convergence criteria for energy and atomic forces were used as
$10^{-6}$ eV and $10^{-5}$ eV/\AA, respectively, in all the
self-consistent-field (SCF) calculations. The Monkhorst Pack
$k$-points of 11$\times$11$\times$1 were used in all calculations.
The rotationally invariant DFT + U approach of Dudarev {\em et al.}
\cite{PhysRevB.57.1505} was used to capture the effect of strongly
correlated 3d, 4d and 5d electrons of V/Nb/Ta in TM-doped MoS$_2$.
Hubbard U parameters were computed self-consistently using density functional
perturbation theory (DFPT), given by Cococcioni {\em et al.}
\cite{PhysRevB.71.035105}. Our computed U values for V, Nb and Ta
are 4.48, 1.18, and 4.49 eV, respectively. For V-doped MoS$_2$, our
calculated U value is in good agreement with the value reported
in the literature \cite{WU2018111}.

To avoid any interaction between layers, a 15 {\AA} vacuum was
incorporated in pristine and doped MoS$_2$ structures along the
$z$-direction, which is reported to be sufficient \cite{yue2013functionalization,
boakye2023electronic}. A 4$\times$4$\times$1 supercell of MoS$_2$ monolayer
was used to incorporate the substitution of transition metals at the Mo site.
The chosen TM concentration for this study is 6.25\%. To examine valley
polarization, we incorporated relativistic effects via spin-orbit coupling
in our calculations. We used the energy-strain method, followed by the
post-processing through VASPKIT \cite{WANG2021108033} to calculate the
elastic coefficients. Further, we calculated piezoelectric coefficients
using density functional perturbation theory (DFPT).
To examine the thermodynamic stability of transition-metal-doped systems,
we performed the \textit{ab initio} molecular dynamics simulations (AIMD) with
a time step of 0.5 fs for 5 ps, at room temperature. We used the Nose
algorithm under the canonical ensemble \cite{10.1063/1.447334} during the
AIMD simulations.

%%%%%%%%%%%%%%%%%%%%%%%% figure 
\begin{figure*}
\includegraphics[width=1.8 \columnwidth,angle=0,clip=true]{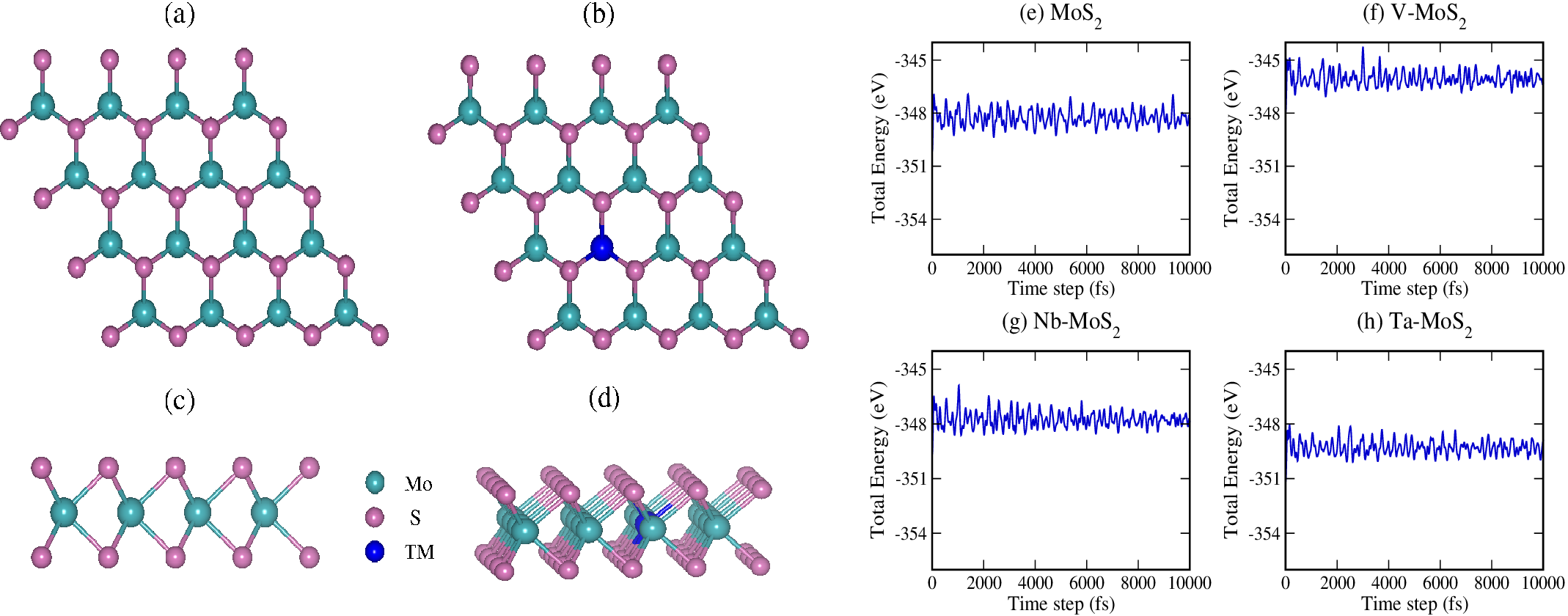}
\caption{Top (a,b) and side (c,d) views of the pristine and TM-doped
        crystals, respectively, of MoS$_2$ monolayer. Transition metal
        elemental substitutional doping considered for TM = V, Nb and Ta.
	Panels (e-h) show the AIMD simulations performed at 300 K for 
	determining the thermal stability.}
	%of (e) monolayer MoS$_2$, (f) V-MoS$_2$, 
	%(g) Nb-MoS$_2$ and (h) Ta-MoS$_2$.}
\label{structure}
\end{figure*}
%%%%%%%%%%%%%%%%%%%%%%%%%%%%%%

%%%%%%%%%%%%%%%%%%%%%%%%%%%%%%%%%
\section{RESULTS AND DISCUSSION}

%%%%%%%%%%%%%%%%%%%%%%%%%%%%%%%%%%%%%%%%%%%%%%%%%%%%%%%%
\subsection{Crystal Structure and Structural Stability}

MoS$_2$ is the well explored material of MX$_2$ (TMDs) family,
in which a layer of transition metal atoms is placed between the
two layers of chalcogen atoms and forms a hexagonal lattice.
It exhibits three forms: 1T, 2H, and 3R, which depend on the
atomic coordination and stacking order. The terms 1T, 2H and 3R
stand for tetragonal, hexagonal, and rhombohedral structures, respectively.
Among these forms, the 2H structure, which is our material of interest,
is thermodynamically stable; however, the 1T and 3R phases are metastable
\cite{santra2024molybdenum}. The point group of bulk 2H-MoS$_2$ is D$_{6h}$,
which possesses inversion symmetry. However, for a monolayer, the crystal
symmetry is reduced to D$_{3h}$. The top and side views of the crystal
structure of pristine MoS$_2$ monolayer are shown in the panels (a) and
(c) of Fig. \ref{structure}.

We begin by optimizing the crystal structure of MoS$_2$ to achieve the
ground state. For this, we start with the experimental structure of MoS$_2$
\cite{PhysRevB.7.3859} and perform full relaxation calculations. The
optimized structural parameters from our calculations are provided in
Table \ref{tab_one}. Our computed lattice parameters for pristine MoS$_2$
are in good agreement with the previously reported lattice parameters
\cite{fang2022van}. As evident from our calculations, TM-substitutions
do not lead to a significant alteration of the lattice parameters. It is
consistent with the previous theoretical study on TM-doped Janus TMDs
WSSe \cite{ZHAO2019172}. For TM-doped systems, the Mo--S bond length differs
slightly, as shown in Table \ref{tab_one}. It could be attributed to the
difference in the atomic size of the Mo atom and the substituted dopant. Top
and side views of the crystal structure of V-doped MoS$_2$ are shown in
panels (b) and (d) of Fig. \ref{structure}.

To investigate the relative stability of these doped structures, we have
examined the binding energies (E$_b$) by using the expression \cite{chen2024electronic}:
\begin{equation}
        E_{b} = E_{\text{doped}} - (E_{\text{Vacancy}} + E_{\text{TM}})
\end{equation}
where, E$_{\text{doped}}$, E$_{\text{Vacancy}}$ and E$_{\text{TM}}$ represent
the total energy of TM-doped MoS$_2$ monolayer, the energy of pristine MoS$_2$
with one Mo vacancy, and the energy of the isolated dopant TM atom,
respectively. The more negative value of E$_b$ means the more
structural stability of the doped system. Table \ref{tab_one}
shows that the computed binding energies are negative for all the
discussed systems. It indicates that doping is energetically favorable,
and the more negative value shows the strong interaction between the
dopant and the parent compound.
Further, to examine the thermodynamic feasibility of these doped systems,
we computed formation energy (E$_F$), which is defined as the energy
required to form the substituted or defect structure. It
is calculated by using this expression \cite{molecules28166122}:
\begin{equation}
E_{f} = [(E_{\text{doped}} - E_{\text{pristine}}) + (\mu_{\text{Mo}} - \mu_{\text{TM}})]
\end{equation}
where, E$_{doped}$ and E$_{pristine}$ are the total energy of the
doped and pristine MoS$_2$, $\mu_{\text{Mo}}$ and $\mu_{\text{TM}}$
are the chemical potentials of the Mo and TM atoms (from the bulk phase).
A small formation energy indicates that the structure is more likely
to form. Our calculated formation energy for V-MoS$_2$ is 4.932 eV, which
is consistent with the reported value of 5.795 eV \cite{boakye2023electronic}.
As shown in table \ref{tab_one}, Nb-MoS$_2$ systems exhibit the lowest
formation energy among all group-5 TM-doped MoS$_2$, which indicates their
high suitability for formation. Nb doping is also reported as a promising
$p$-type dopant in MoS$_2$ monolayer \cite{PhysRevB.88.075420}.

Thermal stability of the layers was investigated by performing AIMD (Ab 
Initio Molecular Dynamics) \cite{molecules28166122} analysis at room 
temperature (300 K). From panels (e-h) of the Fig. \ref{structure},
it is clear that after 10000 AIMD steps, with a time step of 1fs, these doped systems 
remain stable. There was no bond breaking, and the energy fluctuation 
was also very small. Our results demonstrate that the monolayer 
MoS$_2$ and TM-doped MoS$_2$ systems are stable at room temperature.

%%%%%%%%%%%%%%%%%%%%%%%% figure 
\begin{figure*}
\includegraphics[width=1.8 \columnwidth,angle=0,clip=true]{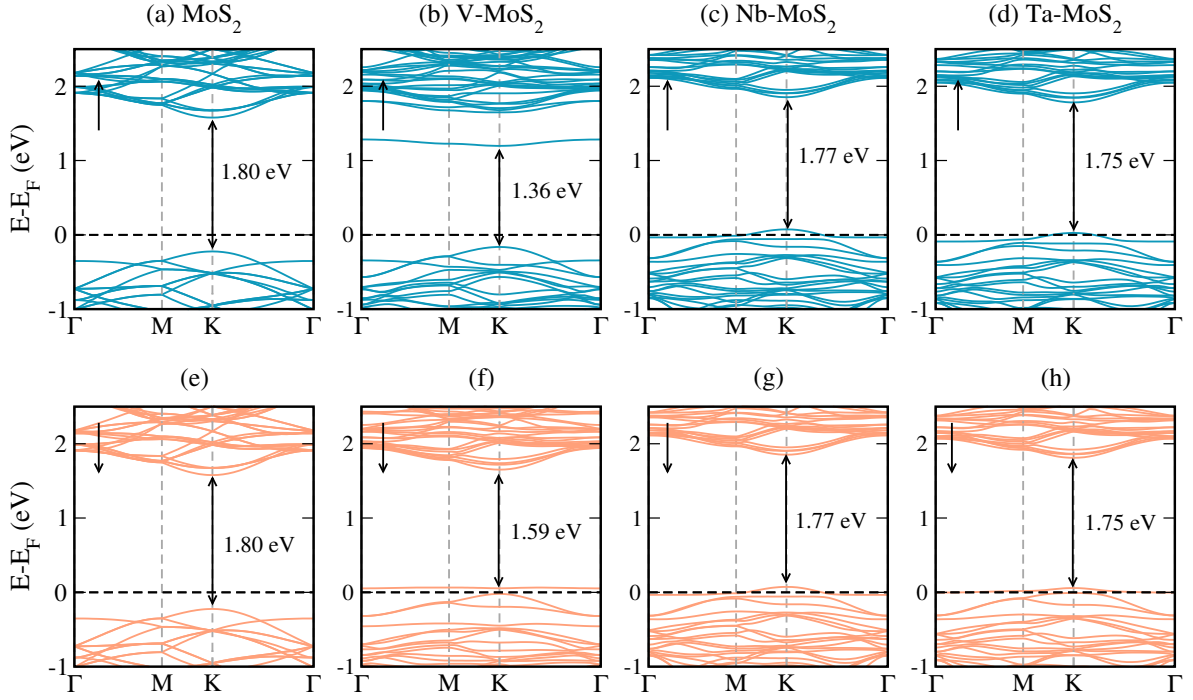}
\caption{Spin-polarized electronic band structures of (a) MoS$_2$, 
	(b) V-MoS$_2$, (c) Nb-MoS$_2$, and (d) Ta-MoS$_2$. The upper 
	and lower panels depict the spin-up and spin-down channels, 
	respectively. The Fermi level is set to zero.}
\label{band}
\end{figure*}
%%%%%%%%%%%%%%%%%%%%%%%%

%%%%%%%%%%%%%% table 1
\begin{table*}
\caption{Calculated lattice parameters, Wyckoff positions, binding
        energies, band gaps, spin polarization, and bond lengths
	and bond angles (in degrees) for TM doped-MoS$_2$ structures.}
\centering
\begin{ruledtabular}
\begin{tabular}{llcccc}
% \hline
&   Lattice constants (\AA)  &  Mo-S (\AA)  & TM-S & S-Mo-S & S-TM-S \\
\hline
MoS$_2$     & 3.15 [Present work] & 2.462 & - & 81.821 & - \\
		    & 3.16 [Theo.] \cite{fang2022van} & & &   &  \\
		    & 3.16 [Exp.] \cite{PhysRevB.7.3859} & & &  & \\
V-MoS$_2$   & 3.15 [Present work]  & 2.400  & 2.353  & 82.971  & 81.853 \\
Nb-MoS$_2$  & 3.15 [Present work]  & 2.411  & 2.446  & 80.997  & 82.366 \\
Ta-MoS$_2$  & 3.15 [Present work]  & 2.414  & 2.444  & 81.014  & 82.362 \\
\hline
        Wyckoff positions &    & $x$  & $y$ & $z$ &    \\
                & Mo  & 0.6667 & 0.3333&  0.7550 & \\
                & S  & 0.3333 & 0.6667 & 0.8067 & \\
\hline
  & Space group & Binding ene.  & Formation ene. (E$_f$) & Bandgap  & Phase  \\
		&      & (eV)    &  (eV)    &  (eV)    &    \\
\hline
 V-MoS$_2$ & 156 & -10.310  & 4.932  & 1.360 ($\uparrow$), 0.074 ($\downarrow$) & Spin gapless semic.    \\

 Nb-MoS$_2$ & 156 & -14.653 & 0.743 & 0.000 ($\uparrow$), 0.000 ($\downarrow$) & Metallic \\

Ta-MoS$_2$ & 156 & -11.756 & 3.640  & 0.000 ($\uparrow$), 0.000 ($\downarrow$)  & Metallic      \\
\end{tabular}
\end{ruledtabular}
\label{tab_one}
\end{table*}

%%%%%%%%%%%%%%%%%%%%%%%%% Fig
%\begin{figure}
%\includegraphics[width=1.0
%\columnwidth,angle=0,clip=true]{aimd.eps}
%\caption{Thermal stability curves as obtained after AIMD simulations at 300 K for 
%	monolayers of (a) MoS$_2$, (b) V-MoS$_2$, 
%	(c) Nb-MoS$_2$ and (d) Ta-MoS$_2$. fs: femtosecond}
%\label{aimd}
%\end{figure}

%%%%%%%%%%%%%%%%%%%%%%%%%%%%%%%%%%%%%%%%%
\subsection{Spin Polarized Band Structure}
%%%%%%%%%%%%%%%%%%%%%%%%%%%%%%%%%%%%%%%%%

Panels (a) and (e) of the Fig. \ref{band} represent the spin-polarized band structures
of the pristine MoS$_2$ monolayer. As is discernible, MoS$_2$ is a direct band gap
semiconductor, with a band gap of 1.80 eV. The predicted electronic structure
from our calculations is in good agreement with the previous reported theoretical
result \cite{fang2022van}.
Then, we investigated the effect of transition metal (TM) (V, Nb and Ta)
-doping on the electronic structure of MoS$_2$.
Panels (b-h) in Fig. \ref{band} represent the spin-polarized band structure
for V/Nb/Ta-MoS$_2$.
The upper and lower panels indicate the spin-up and spin-down channels,
respectively. In the case of V doping (panels (b) and (f)), we observe
a very small band gap in the spin-down channel,
whereas the spin-up channel shows a semiconducting behavior with a band
gap of 1.36 eV. The partially filled 3$d$ orbitals of the V atom introduce
localized impurity or defect states around the Fermi level, which leads
to spin polarization. This mixed nature of the electronic structure
suggests the spin-gapless semiconducting nature of V-MoS$_2$. These types of materials
could be applicable for spintronic applications \cite{liu2024spin}.
On the other hand, both Nb and Ta-doped MoS$_2$ show a state at
Fermi level, as shown in panels (c, d, g and h), which suggests
their metallic nature. However, in the case of Nb-doped MoS$_2$, there
is zero spin polarization. Our computed electronic structure for these
systems are in good agreement with previous reported theoretical results
\cite{yoshimura2020substitutional, li2022controllable, arabinda2025}.

\begin{figure}[h]
\includegraphics[width=1.0
\columnwidth,angle=0,clip=true]{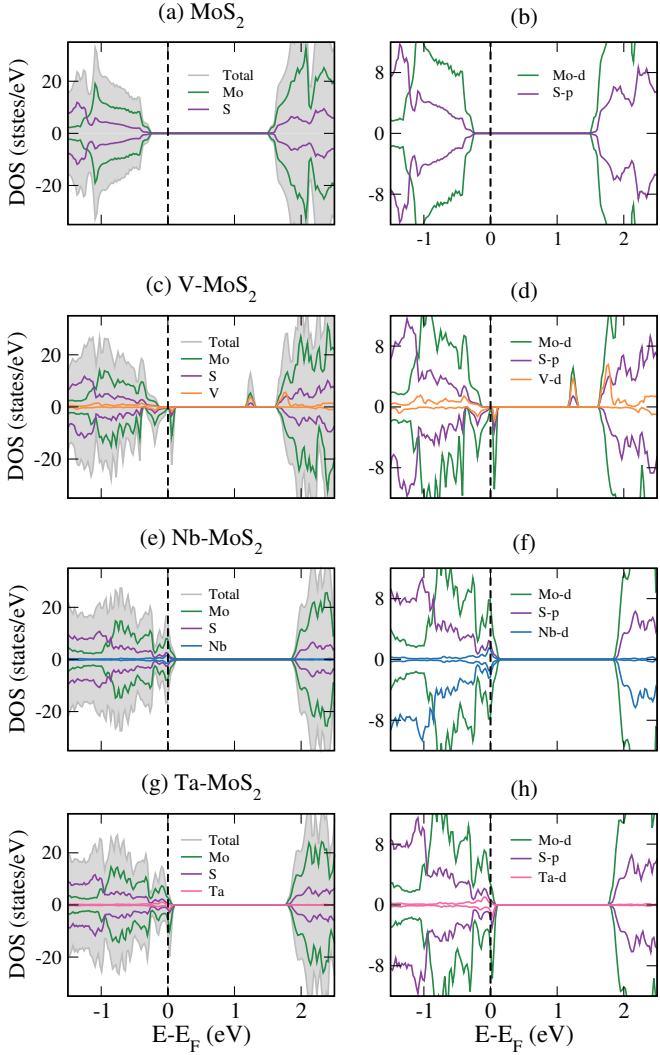}
\caption{Atom- and orbital-projected density of states (PDOS) for pristine and 
	TM–doped MoS$_2$. Panels (a), (c), (e) and (g) show the atom-projected DOS for 
	MoS$_2$, V-MoS$_2$, Nb-MoS$_2$, and Ta-MoS$_2$, respectively, while 
	panels (b), (d), (f) and (h) present the corresponding orbital-projected 
	DOS. The Fermi level is set to zero.}
\label{dos}
\end{figure}

To get further insight into the electronic structure, we examined the
atom and orbital projected density of states (DOS), as shown in Fig. \ref{dos}.
From panels (a) and (b), it is clear that for pristine MoS$_2$,
spin-up and spin-down states are observed to be symmetric, which
indicates an intrinsic nonmagnetic nature. Our computed spin-polarized 
electronic structure also validates it. Here we observed that,
the valence- and conduction-band edges are dominated by 4d orbitals of
Mo and 3p orbitals of S atoms, which is consistent with the reported
previous work \cite{PhysRevB.88.075409}.
For V-doped MoS$_2$, as discernible from panels (c) and
(d), the states near the Fermi level are predominantly contributed
by V-$d$ orbitals with significant hybridization with neighboring
Mo-$d$ and S-$p$ states. Considering the valence electron configurations
of Mo ($4d^{5}5s^{1}$) and V ($3d^{3}4s^{2}$), substitution of Mo by V
introduces a hole in the system. It gives rise to a $p$-type doping and shifts
the Fermi level toward the valence band region, which is consistent with
other reported theoretical literature \cite{miao2020modulation}.
Similar to V doping, both Nb and Ta substitutions act as a p-type
dopant, which shifts the Fermi level near the valence band edges, as
shown in panels (e-h). Although both Nb and Ta doping lead to metallic
character. Our results are in good agreement with previous reported literature
\cite{fan2016ferromagnetism}.

%%%%%%%%%%%%%%%%%%%%%%%% Fig
%%%%%%%%%%%%%%%%%%%%%%%%

Next, we examine the induced magnetism in V/Nb/Ta-MoS$_2$ monolayers.
Previous studies show that the transition metal substitution introduces
magnetism in non-magnetic materials \cite{PhysRevMaterials.9.054003}.
First, we examined the magnetic ground state for all these doped systems.
Therefore, we considered both ferromagnetic (FM) and antiferromagnetic (AFM)
configurations of magnetic moments for V/Nb/Ta-doped systems.
We have computed the relative energy between FM and AFM phases, as given in
Table \ref{tab_mag}. It is clear from the calculated $\Delta E$ that the ground state
magnetic configuration of V and Ta-doped MoS$_2$ is ferromagnetic (FM).
Whereas Nb-MoS$_2$ prefers antiferromagnetic (AFM) as a ground state.
The FM energy is found to be approximately 0.148 and 0.005 eV lower
than from the AFM phase for V and Ta-MoS$_2$, respectively, while
it is higher by 0.009 eV in the case of Nb doping.

Total magnetic moment of pristine MoS$_2$ as well as all doped
systems are given in Table \ref{tab_mag}. As can be expected, our
calculations also predict a zero magnetic moment for pristine MoS$_2$.
In the case of V-MoS$_2$, we observed a total induced magnetic moment of
0.922 $\mu_{\rm B}$, which is primarily due to V atom 0.998 $\mu_{\rm B}$,
while small non-zero magnetic moments are observed on Mo and S atoms
due to $p$-$d$ hybridization. Our computed magnetic moment is in good
agreement with the previous reported theoretical result \cite{boakye2023, MIAO2020109459}.
For Ta, the total induced magnetic moment is 0.624 $\mu_{\rm B}$. Here,
the comparatively small magnetic moment could be attributed to the strong
hybridization between Ta and MoS$_2$ states, which leads to reduced local
magnetic moment of the Ta atom. On the other side, Nb substitution does not
induce magnetism, which is consistent with previous reported literature
\cite{PhysRevB.88.075420}.

%%%%%%%%%%%%%%%%%%%%%%%% table 2
\begin{table}
        \caption{The relative energy ($\Delta E = E_{\rm AFM} - E_{\rm FM}$)
        (in eV) of FM and AFM configurations , the atom resolved magnetic
        moments (in $\mu_{\rm B}/{\rm atom}$) and total magnetic moment $\mu^{\rm Tot}$
	for V/Nb/Ta-substituted MoS$_2$.}
\centering
\begin{ruledtabular}
\begin{tabular}{lrrrrcr}
\% con. &  $\Delta E$ & $\mu^{\rm Mo}$ & $\mu^{\rm S}$ & 
	$\mu^{\rm V/Nb/Ta}$ & $\mu^{\rm Tot}$ \\ \hline
MoS$_2$     & - & - & -  & - & - \\ 
V-MoS$_2$   & 0.148   &  0.005  & -0.005 &  0.998 & 0.922 \\
Nb-MoS$_2$  & -0.009 &  0.000 &  0.000 &  0.000 & 0.000 \\
Ta-MoS$_2$  & 0.005   &  0.031   &  0.002  &   0.094  & 0.624   \\

\end{tabular}
\end{ruledtabular}
\label{tab_mag}
\end{table}
%%%%%%%%%%%%%%%%%%%%%%%%

%%%%%%%%%%%%%%%%%%%%%%%%%%%%%%%%%%%%%%%%%%%%%%%%
\subsection{Valley Polarization}

\begin{figure*}
\includegraphics[width=1.5
\columnwidth,angle=0,clip=true]{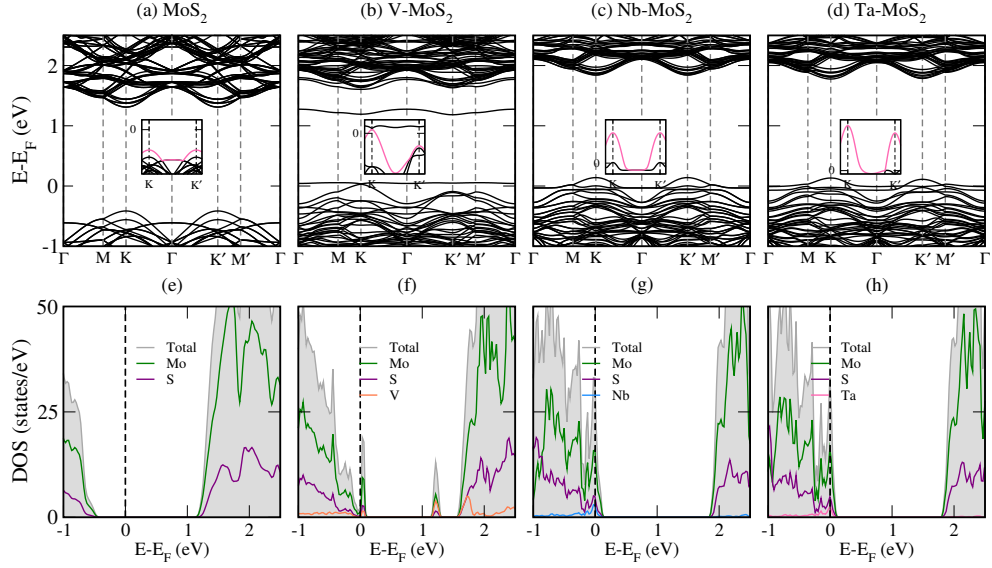}
\caption{The spin-orbit coupled band structures of (a) pristine MoS$_2$
	monolayer, (b) V-MoS$_2$, (c) Nb-MoS$_2$ and (d) Ta-MoS$_2$
	monolayer, respectively. The zoomed views near the Fermi level 
	are shown as insets.}
\label{soc_band}
\end{figure*}

Next, we examined valley polarization in pristine and V/Nb/Ta-doped
MoS$_2$ monolayer. Valley polarization arises from the lifting of
the energy degeneracy between the inequivalent $K$ and $K^\prime$
valleys in the Brillouin zone. It requires the breaking of
time-reversal symmetry and a strong spin-orbit coupling (SOC).
To achieve the valley polarization, the system must break the
time reversal symmetry. The valley polarization is quantified by the energy
difference,
\begin{equation}
\Delta_{KK^\prime} = \left| E_{K^\prime} - E_{K} \right|,
\end{equation}
between the band edges at the $K$ and $K^\prime$ valleys.
Panels (a-d) and (e-h) of the Fig. \ref{soc_band} show the
spin-orbital coupled band structure and density of states of
pristine MoS$_2$, V, Nb and Ta-doped MoS$_2$
monolayer, respectively. It is clear from panel (a), for the pristine case,
that the time-reversal symmetry is preserved, which leads to the same energy
extrema (as shown in the inset of (a)) at the $K$ and $K^\prime$ valleys.
Hence, pristine MoS$_2$ does not exhibit valley polarization. In the case
of TM-doped MoS$_2$ systems, V-doped MoS$_2$ exhibits a pronounced valley
polarization. The induced magnetism in V-MoS$_2$ breaks time-reversal symmetry,
which leads to the difference between the $K$ and $K^\prime$ valleys.
Dopant-induced defect states appear near the Fermi level.
These states are primarily composed due to the hybridization between V--$3d$,
Mo--$4d$, and S--$3p$ orbitals. This hybridization enhances the exchange
interaction and lifts the valley degeneracy, resulting in a valley
polarization of 121 meV, which is consistent with the previously reported
value of 120 meV \cite{singh2017}. There is a flat band above the Fermi level,
which could be due to the additional hole introduced by the V atom.
This feature represents a stable and permanent valley polarization.

It is to be noted that this splitting is comparatively on the higher side
than the obtained splitting of about 0.1-0.2 meV/T, which is produced
by applying an external magnetic field in the MoS$_2$ monolayer
\cite{PhysRevLett.114.037401}.
Chirality-dependent photoexcitation studies further reveal that V-doped
MoS$_{2(1−x)}$Se$_{2x}$ exhibits a room-temperature valley splitting of
approximately 8 meV \cite{Maity_2023}. An even larger splitting has been demonstrated in
monolayer WS$_2$ monolayer, where a giant valley exciton splitting of
16 meV/T arises due to the magnetic proximity effect from an EuS
substrate \cite{norden2019giant}. More advanced approaches based on magnetic proximity effects
in van der Waals heterostructures, have pushed this limit further; for
instance, WSe$_2$/CrI$_3$ heterostructure exhibits an experimentally
measured valley splitting of 34.01 meV, corresponding to an effective
magnetic field of 170 T \cite{ZHANG2024158986}.
Therefore, V-doping offers a robust platform for
valleytronic-based applications without external excitation.
In contrast, Nb-doped MoS$_2$ does not show valley polarization, as
shown in panel (c), which is due to its non-magnetic nature.
For Ta-doped MoS$_2$, a finite but relatively very small valley
polarization of approximately 21 meV is observed.
Although the Ta atom enhances the spin-orbit coupling effect, the induced
magnetism is weak compared to V-doped MoS$_2$, which leads to small
valley polarization.

%%%%%%%%%%%%%%%%%%%%%%%%% fig

%%%%%%%%%%%%%%%%%%%%%%%%%%%%%%%%%%%%%%%%%%%%%%%%%%
\subsection{Piezoelectric Properties}
%%%%%%%%%%%%%%%%%%%%%%%%%%%%%%%%%%%%%%%%%%%%%%%%%

In order to investigate the piezoelectric response of the V-, Nb-,
and Ta-doped MoS$_2$ monolayers, we first examined their mechanical
stability and elastic behavior. Piezoelectricity arises due to the
coupling between mechanical deformation and polarization, therefore,
elastic constants play a fundamental role in exploring the piezoelectric
response \cite{duerloo2012intrinsic}. We have systematically examined
the elastic constants C$_{ij}$ (i, j = 1, 2, ... 6) using the energy-strain
method \cite{WANG2021108033}. To ensure the mechanical stability of the
discussed systems, the calculated elastic constants were examined under the
Born stability criteria. According to which, the necessary and sufficient
condition of elastic stability for the hexagonal system is given by these
equations \cite{PhysRevB.90.224104},
\begin{equation}
\begin{aligned}
C_{11} &> |C_{12}|, \\
2C_{13}^{2} &< C_{33}\left(C_{11}+C_{12}\right), \\
C_{44} &> 0, \\
C_{66} &> 0 .
\end{aligned} 
\end{equation}
However, as in the case of a 2D hexagonal system the out-of-plane
components are zero, only $C_{11} (= C_{22})$, $C_{12}$, and $C_{66}$ 
are non-zero parameters. And among them, $C_{11}$ and $C_{12}$ are the 
only two independent parameters. Parameter $C_{66}$ can be expressed 
in terms of $C_{11}$ and $C_{12}$ as
\begin{equation}
C_{66} = \frac{1}{2}\left(C_{11} - C_{12}\right).
\end{equation}
Similarly, the other mechanical parameters such as Young's 
modulus ($Y$) and shear modulus ($G$) can be derived 
using $C_{11}$, $C_{12}$, $C_{66}$ as
\begin{align}
  Y &= \frac{C_{11}^2 - C_{12}^2}{C_{11}}, \\
  G &= C_{66}.
\end{align}

%%%%%%%%%%%%%%%%%%%% table
\begin{table*}
\centering
\caption{Elastic constants ($C_{11}$ and $C_{12}$), Young's modulus and 
	shear modulus (in $(\mathrm{N/m})$), and piezoelectric coefficients $e_{11}$ 
	$(10^{-10}\,\mathrm{C/m})$ and $d_{11}$ $(\mathrm{pm/V})$ for all 
	studied systems. Both clamped-ion and relaxed-ion contributions 
	are included.}
\label{tab_three}
%\small
\begin{ruledtabular}
\begin{tabular}{lcccc|cccc|cc}
%\hline
& \multicolumn{4}{c}{Clamped-ion} & \multicolumn{4}{c}{Relaxed-ion} \\
\hline
System & $C_{11}$ & $C_{12}$ & $e_{11}$ & $d_{11}$ & $C_{11}$ 
	& $C_{12}$ & $e_{11}$ & $d_{11}$ & Y & G \\
	& $(\mathrm{N/m})$ & $(\mathrm{N/m})$ & $(10^{-10}\,\mathrm{C/m})$ & 
	$(\mathrm{pm/V}) $ & $(\mathrm{N/m})$ & $(\mathrm{N/m})$ & $(10^{-10}\,\mathrm{C/m})$ 
	& $(\mathrm{pm/V})$ & $(\mathrm{N/m})$ & $(\mathrm{N/m})$ \\
\hline\
MoS$_2$ & 168.45 & 49.54 & 3.04 & 2.56 & 144.91 
	& 33.37 & 3.63  & 3.25 & 137.23  & 55.77  \\
	& 153 \cite{duerloo2012intrinsic} & 48 \cite{duerloo2012intrinsic} 
	& 3.06 \cite{duerloo2012intrinsic} & 2.91 \cite{duerloo2012intrinsic} & 
	130 \cite{duerloo2012intrinsic} & 32 \cite{duerloo2012intrinsic} & 3.64 
	\cite{duerloo2012intrinsic} & 3.73 \cite{duerloo2012intrinsic} &  & \\
V-MoS$_2$  & 162.14 & 45.37 & 6.10 & 5.23 & 129.48 
	   & 21.55 & 6.33 & 5.87 & 125.90 & 53.96  \\
Nb-MoS$_2$ & 165.19 & 45.70 & 4.52 & 3.78 & 130.30 
	   & 20.67 & 4.78 & 4.36 & 127.02 & 54.81 \\
Ta-MoS$_2$ & 167.26 & 46.31 & 4.29 & 3.55 & 132.85 & 21.92 & 4.47 & 4.03 & 129.23 & 55.46 \\
\end{tabular}
\end{ruledtabular}
\end{table*}
%%%%%%%%%%%%%%%%%%%%%%%%%%%%%%%%%

\begin{figure}[h!]
\includegraphics[width=0.8
\columnwidth,angle=0,clip=true]{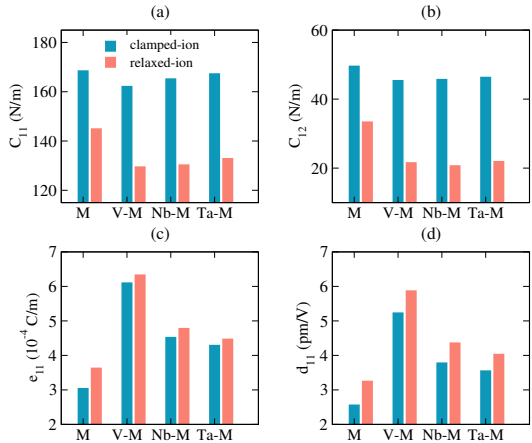}
\caption{Panels (a) and (b) represents the elastic coefficients C$_{11}$ and C$_{12}$,
        and Panels (c,d) represents the piezoelectric coefficients d$_{11}$ and e$_{11}$ 
	of pristine and doped monolayer systems. Here, M, V-M, Nb-M, Ta-M denote 
	pristine, V-doped, Nb-doped, and Ta-doped MoS$_2$ monolayers, respectively.}
\label{piezo}
\end{figure}

In the present work, the total energy is calculated with strain
components ranging from $-0.006$ to $0.006$ in steps of $0.002$.
To calculate the relaxed-ion coefficients, the atomic positions are
fully relaxed at each strain value. In contrast, when the
calculation is performed at each strain without allowing
atomic relaxation, the resulting coefficients are referred
to as clamped-ion coefficients. However, the experimentally observed
quantities are relaxed-ion coefficients \cite{duerloo2012intrinsic}.
Clamped- and relaxed-ion elastic coefficients for all considered
systems are summarized in Table \ref{tab_three}.
It is clear from the table that our studied systems fulfil
the Born stability criteria, which confirms their
mechanical stability. Our computed elastic coefficients for the MoS$_2$ monolayer
are in good agreement with the available theoretical \cite{duerloo2012intrinsic} results.
As shown in panels (a) and (b) of Fig. \ref{piezo}, for the TM-doped
systems, a reduction in the elastic constants is observed compared to
pristine MoS$_2$. Among the doped systems, V-doped MoS$_2$ exhibits the
largest reduction in the elastic constants $C_{11}$ and $C_{12}$. However,
Nb and Ta-doped MoS$_2$ retain relatively higher values of both elastic
constants. These results demonstrate that transition-metal doping leads
to a modulation of the in-plane elastic behavior of MoS$_2$. These
variations in $C_{11}$ and $C_{12}$ are expected to directly affect
the electromechanical coupling in the doped monolayers, which is
related to the piezoelectric properties.

Next, we investigate the piezoelectric coefficients by using density functional
perturbation theory (DFPT). The piezoelectric effect is a first-order coupling
between polarization ($P_i$) or the electric field ($E_i$), and stress
($\sigma_{jk}$) or the strain ($\varepsilon_{jk}$) tensors, where $i, j, 
k \in \{1,2,3\}$, with 1, 2, and 3 correspond to $x$, $y$, and $z$,
respectively. Therefore, the piezoelectric tensors $d_{ijk}$ and $e_{ijk}$ 
can be expressed as
\begin{equation}
d_{ijk} = \left( \frac{\partial P_i}{\partial \sigma_{jk}} \right)_{E,T} 
	= \left( \frac{\partial \varepsilon_{jk}}{\partial E_i} \right)_{\sigma,T} {\rm and}
\end{equation}
\begin{equation}
e_{ijk} = \left( \frac{\partial P_i}{\partial \varepsilon_{jk}} \right)_{E,T} 
	= -\left( \frac{\partial \sigma_{jk}}{\partial E_i} \right)_{\varepsilon,T}
\end{equation}
It is to be noted that monolayer 2H-MoS$_2$ has $D_{3h}$ ($6m2$) point 
group, which restricts all coefficients to be zero except $d_{11}$ and $e_{11}$.
Piezoelectric stress and strain coefficients $d_{11}$ and $e_{11}$, respectively, 
are related through elastic coefficients by the relation
\begin{equation}
        d_{11} = e_{11}/ {(C_{11} - C_{12})}
\end{equation}

%%%%%%%%%%%%%%%%%%%%%%%%% figure 

Our calculated clamped-ion and relaxed-ion piezoelectric coefficients 
$d_{11}$ and $e_{11}$ for all considered systems are given in Table 
\ref{tab_three}. It is clear from the panels (c) and (d) of Fig. 
\ref{piezo}, that the doping of V, Nb and Ta enhances the $d_{11}$ 
and $e_{11}$ coefficients. It is to be noted that the enhancement 
in the piezoelectric coefficients in TM-doped MoS$_2$ systems could 
be explained in terms of bond length, atomic size and electronic
structure. The V-S bond length (2.353 ) is shorter than the Mo-S band 
(2.462 ), which indicates the strong bonding and enhanced charge 
redistribution. However, in the case of Nb and Ta, they show a smaller
change in bond length, which leads to relatively weaker structural 
contribution to polarization. This stronger bonding promotes a larger 
internal strain contribution to the piezoelectric coefficient. Moreover, 
the smaller atomic radius of V, compared to Nb and Ta, induces a more 
compact and locally stabilized bonding environment without introducing 
excessive lattice distortion, thereby preserving the non-centrosymmetric
symmetry essential for piezoelectric polarization. Additionally, the
more diffuse 4d (Nb) and 5d (Ta) orbitals enhance electronic screening
effects, which can partially reduce the effective polarization response
under strain. The combined structural and electronic factors, therefore,
explain the observed trend in piezoelectric performance. Our results
are consistent with a recent experimental study on Nb-doped MoS$_2$
based piezoelectric nanogenerators \cite{dhakar2024significant}, where
the output voltage was enhanced by approx. 2.5 times compared to pristine
MoS$_2$ monolayer. In our calculations, the enhancement factor is
approximately. 1.47 times, which confirms that the substitutional doping
enhances the piezoelectric response.

%\begin{table}
%\caption{ The elastic constants (C$_{11}$ and C$_{12}$) 
%	in units of N/m, Young's and shear modulus in N/m, 
%	piezoelectric coefficient (e$_{11}$) in pC/m and d$_{11}$ 
%	in pm/V for all discussed systems.}
%\centering
%\begin{ruledtabular}
%\begin{tabular}{ccccccc}
%System  & C$_{11}$ & C$_{12}$ & Y & G  & e$_{11}$ & d$_{11}$ \\
%        & (N/m) & (N/m) & (N/m) & (N/m) & (pC/m) & (pm/V) \\ 
%\hline\\
%MoS$_2$    & 168.45 & 49.54 & 153.88  & 59.45  & 304.82  & 2.56  \\
%           & 153 \cite{duerloo2012intrinsic} & 48 \cite{duerloo2012intrinsic} & & & 306 \cite{duerloo2012intrinsic}  &  2.91 \cite{duerloo2012intrinsic} \\   
%V-MoS$_2$  & 162.14 & 45.37 & 149.44  & 58.38  & 610.85  & 5.23  \\
%Nb-MoS$_2$ & 165.19 & 45.70 & 152.55  & 59.74  & 452.43  & 3.78  \\
%Ta-MoS$_2$ & 167.26 & 46.31 & 154.44  & 60.47  & 429.63  & 3.55  \\
%\end{tabular}
%\end{ruledtabular}
%\label{tab_three}
%\end{table}

%%%%%%%%%%%%%%%%%%%%%%%%%%%%%%%%%%%%%%%%%%%%
\section{Conclusions}

In summary, with the help of density functional theory-based first-principles
calculations, we have investigated the effect of group-5 transition-metal
(V, Nb, and Ta) substitution on the electronic, magnetic, piezoelectric,
and valleytronic properties of monolayer MoS$_2$. From our study, we found
the emergence of spin-gapless semiconducting nature for V-doped MoS$_2$,
which could offer a robust platform for spin transport applications. On
the other hand, both Nb and Ta-doped MoS$_2$ exhibit metallic behavior,
which occurs due to impurity-induced defect states at the Fermi level.
Our calculated total induced magnetic moment for V- and Ta-doped MoS$_2$,
is 0.922 and 0.624 $\mu{\rm B}$, respectively.
Apart from spin polarization, our study on valley polarization predicts
the values 121 and 21 meV for V and Ta substitutions, respectively. More
importantly, our study also suggests the enhanced piezoelectric response for
V/Nb/Ta-doped systems, compared to the pristine MoS$_2$ monolayer.
The coexistence of spin-polarization, significant valley polarization,
and enhanced piezoelectricity suggests group-5 TM-MoS$_2$ monolayers
could offer a potential multifunctional material for spintronic,
valleytronic and piezoelectric devices.

%%%%%%%%%%%%%%%%%%%%%%%%%%%%%%%%%%%%%%%%%%%%
\begin{acknowledgments}
BKM acknowledges the funding support from SERB, DST (CRG/2022/000178).
Shivani acknowledges the fellowship support from UGC (BININ01949131), 
Govt. of India.
The calculations are performed using the High Performance Computing cluster Tejas
at the Indian Institute of Technology Delhi and PARAM Rudra, a national supercomputing
facility at Inter-University Accelerator Centre (IUAC) New Delhi.
\end{acknowledgments}

%%%%%%%%%%%%%%%%%%%%%%%%%%%%%%%%%%%%%%%%%%%%%%%%%%%%%%%%%%%%%%%%%%%%%%%%%%%%%%%%%%%%%%%%%%
%%%%%%%%%%%%%%%%%%%%           References        %%%%%%%%%%%%%%%%%%%%%%%%%%%%%%%%%%%%%%%%%
%%%%%%%%%%%%%%%%%%%%%%%%%%%%%%%%%%%%%%%%%%%%%%%%%%%%%%%%%%%%%%%%%%%%%%%%%%%%%%%%%%%%%%%%%%
\newpage

\bibliography{reference}
\end{document}